# Efficient Forward Second-Harmonic Generation from Planar Archimedean Nanospirals

*Roderick B. Davidson II[1,*], Jed I. Ziegler[1], Guillermo Vargas[1,†], Sergey M. Avanesyan[1], Richard F. Haglund Jr.[1]*

*Yu Gong[2], Wayne Hess[2]*

[1]Department of Physics and Astronomy, Vanderbilt University, Nashville, TN, USA

[2]Physical Sciences Division, Pacific Northwest National Laboratory, Richland, WA, USA

Abstract: The enhanced electric field at plasmonic resonances in nanoscale antennas can lead to efficient harmonic generation, especially when the plasmonic geometry is asymmetric on either inter-particle or intra-particle levels. The planar Archimedean nanospiral offers a unique geometrical asymmetry for second-harmonic generation (SHG) because the SHG results neither from arranging centrosymmetric nanoparticles in asymmetric groupings, nor from non-centrosymmetric nanoparticles that retain a local axis of symmetry. Here we report forward SHG from planar arrays of Archimedean nanospirals using 15 fs pulse from a Ti:sapphire oscillator tuned to 800 nm wavelength. The measured harmonic-generation efficiencies are $2.6 \cdot 10^{-9}$, $8 \cdot 10^{-9}$ and $1.3 \cdot 10^{-8}$ for left-handed circular, linear, and right-handed circular polarizations, respectively.

The uncoated nanospirals are stable under average power loading of as much as 300 µW per nanoparticle. The nanospirals also exhibit a selective conversion between polarization states. These experiments show that the intrinsic asymmetry of the nanospirals results in a highly efficient, two-dimensional harmonic generator that can be incorporated into metasurface optics.

The second-order susceptibility governs a host of important nonlinear optical phenomena, including frequency mixing, sum-frequency and harmonic generation, and optical rectification. In crystalline and molecular materials, second-order nonlinearities are non-vanishing only at surfaces or in materials with a non-centrosymmetric crystal structure; moreover, the efficient generation of a second-order nonlinear effect requires that the fundamental incident and the nonlinear outgoing waves be phase matched through a macroscopic volume of material, typically on the order of cubic millimeters.[1]

Plasmonic nanostructures and nanostructure arrays also exhibit second-order nonlinearities, and can generate forward second harmonics if their geometries are not centrosymmetric. Such structures are inherently planar, and therefore compatible with thin-film optical and opto-electronic technologies and with metasurface optics. With advances in nanofabrication, the symmetry of the structures can be exquisitely controlled at the level of a few nanometers.[2] Combining these effects with ultrafast, high-intensity laser pulses yields massive electric-field enhancements and correspondingly greater second-harmonic yield. The localized surface plasmon resonance enhances efficiency, and can be designed by selecting the nanoparticle shape for a given wavelength; selective polarization response can also be designed into the nanoparticle. A number of asymmetric plasmonic geometries have been used for harmonic generation, including L- and V-shaped nanoparticles, nanocups and asymmetric trimers.[3-7]

Larger plasmonic structures – such as the ratchet wheel – have also been shown to affect the polarization of harmonic emission.[8-12]

However, even though metallic nanostructures can generate significant second-order nonlinear responses per volume, the nanostructure volume is incredibly small.[13] High efficiency is therefore also crucial in order to generate reasonable numbers of photons in nonlinear processes without melting or damaging the nanostructures. Aperiodic arrangements of centrosymmetric nanoparticles offers one route to enhanced nonlinear signals.[14] Creating non-centrosymmetric systems of particles with a center of inversion symmetry yields even greater harmonic conversion efficiencies.[15] The need for efficient nanostructured harmonic generators then drives the search for new plasmonic geometries with higher harmonic generation efficiencies in spite of low material volumes.

In this paper, we describe the second-order nonlinear response from arrays of planar 4π Archimedean nanospirals with sub-wavelength dimensions. The Archimedean nanospiral commends itself as a frequency-conversion architecture due to its unique asymmetry and two-dimensional chiral response. Previous experiments and simulations have shown that this geometry has a spectrally complex response in the visible to the near-infrared region and spatially differentiated, near-field configurations, as well as selectively enhanced optical response to the polarization states of incident light.[16-17] These characteristics make the nanospiral a strong candidate for nonlinear optical applications where a broadband plasmonic element is necessary. Unlike plasmonic structures with globally broken symmetry created by modifying or arranging nanoparticles with some inherent local symmetry, [3-4, 7, 14] the nanospiral has no local axes of symmetry at all so that the nanospiral can generate second-harmonic light from any polarization state. This inherent lack of symmetry therefore makes the nanospiral an attractive candidate for nonlinear metasurface elements.

Figure 1(a) shows the experimental arrangement in which SHG signals from nanospiral arrays were measured. The illumination source for these experiments was a Ti:sapphire oscillator (KM Labs Cascade) with an output spectrum centered at 800 nm. The oscillator beam was directed through a 128-pixel, double-mask, spatial light modulator (SLM, Biophotonics Solutions) that uses multiphoton intrapulse interference phase scanning (MIIPS) to compress the 50 fs oscillator pulse to a transform-limited duration of 15 fs.[18-19] The laser pulse was focused on to the nanospiral array using a lens with a numerical aperture of 0.35 to create a focal spot size of 10μm. The maximum energy per pulse was 0.33 pJ at a repetition rate of 82 MHz and was varied using a half-wave plate and linear polarizer combination. After passing through the nanospiral array, the fundamental (800nm) was filtered out of the signal using a short-pass filter centered at 625 nm and a band-pass filter centered at 400 nm. The SHG signal produced by the nanospirals was detected using a solid-state photomultiplier tube (Hamamatsu, RU-9880U-110) in connection with a photon-counting system (Stanford Research Systems).

The nanospirals used in these experiments were created using electron beam lithography in a Raith eLINE scanning electron microscope. All experiments demonstrated here used 4π rotation nanospirals ordered in 10μm x 10μm arrays on ITO-covered glass substrates. The Archimedean nanospiral is described by the equation $r = \alpha\theta$, where α is a constant that determines the rate of expansion of the spiral. These particular spirals were designed such that the spacing between each subsequent arm is equal to the others. The inter-particle spacing was 610nm with a maximum nanoparticle diameter of 395nm and a thickness of 40nm

Figure 1(b) shows the measured second-harmonic conversion efficiency from the nanospiral array (red triangles) as a function of the incident laser power at 800 nm with linear polarization. The data are fit to a second-order polynomial computed in Matlab® (continuous blue curve) with an $R^2$ value of 0.9911 as calculated. Extensive computer simulations of the spectral distribution of near-field modes of the nanospiral show that the dominant spatial configuration of the plasmon near 800 nm wavelength is the focusing mode, in which the near-field electromagnetic energy is concentrated in the center of the nanospiral.[16,17] In order to see if indeed the focusing mode is the dominant source of the SHG signal, the nanospirals were observed under a photo-emission electron microscope (PEEM). The top left inset of Figure 1(b) shows an intensity map of photo-emitted electrons from nanospirals being excited simultaneously with a white-light neon lamp and an 800nm ultrafast laser pulse with a measured duration of order 50 fs. The white light source reveals the overall contours of the nanospiral while the ultrafast pulse creates intense areas of emission that correspond to the electric-field intensity generated in the plasmon near field. Since PEEM is a two-photon process that depends on the electric-field of the plasmon in the same way as SHG, this verifies that most of the observed SHG signal is coming from the center of the nanospiral. This makes the focusing mode ideal for harmonic generation, since second-order nonlinear phenomena depend superlinearly on the strength of the electric field. The focusing mode of the nanospiral provides the spatial concentration of optical energy necessary for efficient conversion to the outgoing second harmonic. The structure of this near-field state also provides a polarization-sensitive template for quasi-chiral properties that are described later in this paper.

The SHG measurements were not reproducible at powers greater than 280 µW per nanoparticle because laser heating deformed the nanospirals, causing them to lose their asymmetric geometry;

the shape of the deformations that occurred due to melting can be seen in the Supplemental Material, Figure S1. Up until the point of deformation, the second-harmonic beam from the nanospirals showed no sign of saturation due to nonlinear down-conversion processes. After deformation, the nanoparticles exhibit neither second-harmonic response nor the other properties of the nanospirals. If the nanospirals were coated in a protective silicon layer as shown in references 3 and 7, the power threshold could almost certainly be increased beyond an incident power of 280 µW per nanoparticle in order to create even higher harmonic-generation efficiencies.

Fig. 2(a) shows that the SHG signal is modulated in angle as the axis of linear polarization is rotated with respect to the nanospiral axis. This change in signal corresponds to the changes in the computed near-field electric-field strengths in the center of the nanospiral shown in Figure 2(b-e). The maximum SHG signal occurs along the 30˚-210˚ axis where the peak electric-field amplitude of the plasmon is the strongest; the field strength is distributed largely within the interior of the spiral and follows the rotation of the spiral arm through $\pi$ radians [Fig. 2)d)]. The maximum field strength is higher than for other orientations, and the dipoles excited near the center are spatially coherent. Other orientations of the linear polarization vector correspond to decreasing SHG conversion efficiency due to a dissipation of the strong focusing mode along the exterior regions of the spiral. The electron concentrations are spread out well away from the center, and also create incoherent multipolar resonances throughout the spiral. The second-harmonic radiation produced by these physically separate resonances destructively interferes with itself as expected from SHG produced by a spatially symmetric set of electronic dipole resonators.[14] These symmetry effects are evident from brighter edges of the spiral in the simulations.

The second-harmonic response to circular polarizations, however, more clearly illustrates the effects of intra-particle resonances on the electric field strengths and consequently the SHG efficiency. The polarization dependence of the SHG intensity in Fig. 3 shows that for right-handed circular polarization – rotating from the outside of the spiral to the inside - there is a larger enhancement than that observed with linear polarization. When excited with left-handed circularly polarized light the nanospirals show a significantly reduced second-order response. The maximum and minimum SHG signals differ by a factor four, which occurs for an eccentricity of 0.66, where eccentricity is defined as the ratio of minor to major axis of the ellipse traced out by the polarization vector. This dependence on eccentricity arises because the nanospiral is not perfectly circular. The second harmonic conversion efficiency – given by $I(2\omega)/I(\omega)$ - is thus a maximum for right circular polarized light and has the value $1.3 \cdot 10^{-8}$.

The nonlinear polarizability that is responsible for second order phenomenon can be written in this case as $P(2\omega) = 2\varepsilon_0 d_{eff}[E(\omega)]^2$ where $d_{eff}$ is the effective second order susceptibility (normalized to the measured efficiency) and E(ω) is the fundamental pump field. Using a driven Lorentz oscillator model appropriate for electrons in gold, we find that $d_{eff}$ ranges from 15.3 pm/V for RCP excitation to 3 pm/V for LCP excitation. These values are comparable to the $d_{eff}$ measured for tilted gold hemispherical nanoparticles.[7] Even though the nanospiral extinction varies as a function of the incident polarization state, there is nevertheless a finite SHG response to all polarizations. This novel property of the nanospiral is a consequence of the absence of any local axis of symmetry.

The polarization response of the second harmonic emission is consistent with FDTD simulations that have been performed with this plasmonic geometry in previous work.[16-17] Simulations using Lumerical Solutions® software show three unique near-field structures in the plasmon resonances that occur at spectrally distinct points in the optical and near-infrared band, and have been designated as standing wave, hourglass, and focusing modes. At a wavelength of 800nm, the focusing mode is excited, concentrating the near-field intensity in the center of the spiral to create a single region of high electric field.[16] Right-handed circularly polarized light assists this electron concentration by driving the electrons from the outer regions of the nanoparticle towards the center. This, in turn, generates increased electric field amplitude and consequently a larger SHG signal. Left-handed circular polarization causes the opposite effect by driving the electrons away from the center of the spiral.

Three-dimensional chiral geometries exhibit strong responses to polarization and a geometry that cannot be superimposed upon itself as discussed, for example, in Refs. 21-22. The existence of two-dimensional chirality in plasmonic structures has sometimes been debated, because even thin nanostructures – such as the nanospirals in our experiment, which have an aspect ratio of order 16 – are not rigorously planar. Nevertheless, quasi-two-dimensional plasmonic nanostructures with chiral boundary conditions in only two dimensions have been shown to rotate linear polarized light and to convert one polarization state to another.[23-25] This property can be identified as one – albeit only one – characteristic of a chiral system.

We now show that the second harmonic generated by Archimedean nanospirals reveals the complex interplay among polarization states that is the hallmark of this quasi-chiral response. The conversion between linear and circular polarizations was investigated by placing a second quarter-wave plate and linear polarizer in the path of the SHG emission. Figure 4 (a-c) show the polarization profiles for three types of incident polarization using only a linear polarizer as an analyzer. The dipolar pattern in Figure 4 (a) shows that when the nanospirals are illuminated by linearly polarized light, the SHG emission is linearly polarized about the same axis as the exciting fundamental beam. This is not surprising, because the localized dipolar plasmon resonance responds to all polarizations. When the fundamental beam is right-circularly polarized, as in Figure 4 (b), the SHG emission is still linearly polarized, but the SHG axis of polarization is pinned on the axis 135 degrees from horizontal. Lastly, a fundamental beam with left-handed circular polarization, Figure 4 (c), does not produce linearly polarized SHG emission at all.

However, the left-handed circular polarization experiment in Figure 4 (c) only shows that the emitted light is not linearly polarized; this result could have occurred either because the SHG was circularly polarized or because it was essentially depolarized. To resolve this question, a quarter-wave plate was inserted between the nanospirals and the linear analyzer, which would have converted any circularly polarized light to the linearly polarized state. Since no axis of linear polarization was evident in the analyzed SHG signal, it was clear that left-handed circular polarized light incident on the nanospirals creates depolarized second-harmonic light. While the signal is lower than SHG from linear or RCP light, it is only a factor of four lower than the maximum signal observed from the nanospirals.

This behavior seems to be reproduced by the focusing mode simulations shown in Figure 5. When the polarization is driving the electrons towards the center of the nanospiral as shown in Figure 5 (a), the strength of the electric field is enhanced. The surface charge density of the plasmon must be ordered if it is to maintain the constructively interfering, multipolar resonance condition that will prevent second-order light from destructively interfering with itself.

When the polarization is driving the electrons away from the center of the nanospiral, on the other hand, a disordered plasmon resonance is created that is simulated in Figure 5 (b). Depolarization in plasmon emission has been observed in simple nanoparticle geometries having resonances that overlap spectrally and spatially; the coherent emission from sub-wavelength structures may exhibit partial depolarization in various directions depending on the relative phases of the light.[26-27] The near-field structure of the LCP excitation in the nanospiral shows regions of high-electric field enhancement that are spatially spread out over larger parts of the nanospiral than in either RCP or linear excitation. This structure contains enough smaller resonances with randomly oriented propagation directions that the resulting SHG emission is completely depolarized. While the polarization conversion observed in nanospirals differs from the polarization rotation that has been demonstrated in other plasmonic geometries[23-25], the conversion between linear, circular, and depolarized light is evidence of the complex interplay between the second-order response and the near-field structure of the plasmon resonance.

In summary, the Archimedean nanospiral is shown here to produce second-harmonic emission at intensities sufficient for nanotechnology devices. The capacity to modulate the intensity of the SHG by altering the polarization state of the emission, and with efficiencies as large as $1.3 \cdot 10^{-8}$ at 280 µW incident power per nanoparticle, the nanospiral is a competitive architecture for all-optical control applications. The nonlinear response to the handedness of circular polarization also reveals the relation between the complex boundary conditions imposed by the spiral shape of the nanoparticle and the plasmon resonance.

Planar frequency-conversion structures are an essential element of plasmonic circuitry.[28] By creating more efficient SHG structures that need not satisfy any phase matching conditions, and by using localized surface plasmon (LSP) resonances to further enhance efficiency, we have shown that the nanospiral is potentially significant addition to plasmonic technology. The complex plasmonic resonance structure inherent in the nanospiral, and its complex, but selectable, polarization response, bring additional dimensions to the search for efficient nonlinear plasmonic light sources.

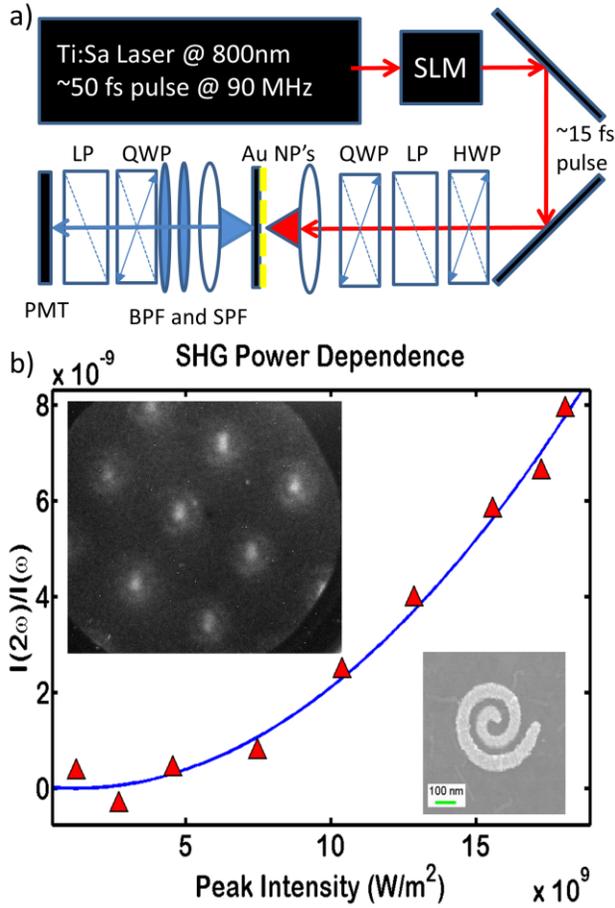

Figure 1 a) Experimental setup used for SHG detection. QWP – quarter wave plate. LP – linear polarizer. HWP – half-wave plate. BPF – band pass filter. SPF – short pass filter. SLM – spatial light modulator. b) FDTD simulation of the nanospiral's near-field structure (focusing mode) of the plasmon resonance excited at 800nm wavelength. Inset) SEM image of a nanospiral with 4pi rotation, used for all experiments presented. c) Power dependence of SHG signal on incident power with linear polarization. Blue line – second order polynomial curve fit. d) Angular diffraction pattern from nanoparticle array. Blue line- polarization axis aligned with smallest width of the nanospiral. Green line – polarization axis aligned with largest width of the nanospiral. Inset – Blue line – SHG spectrum Red line – Fundamental pulse spectrum.

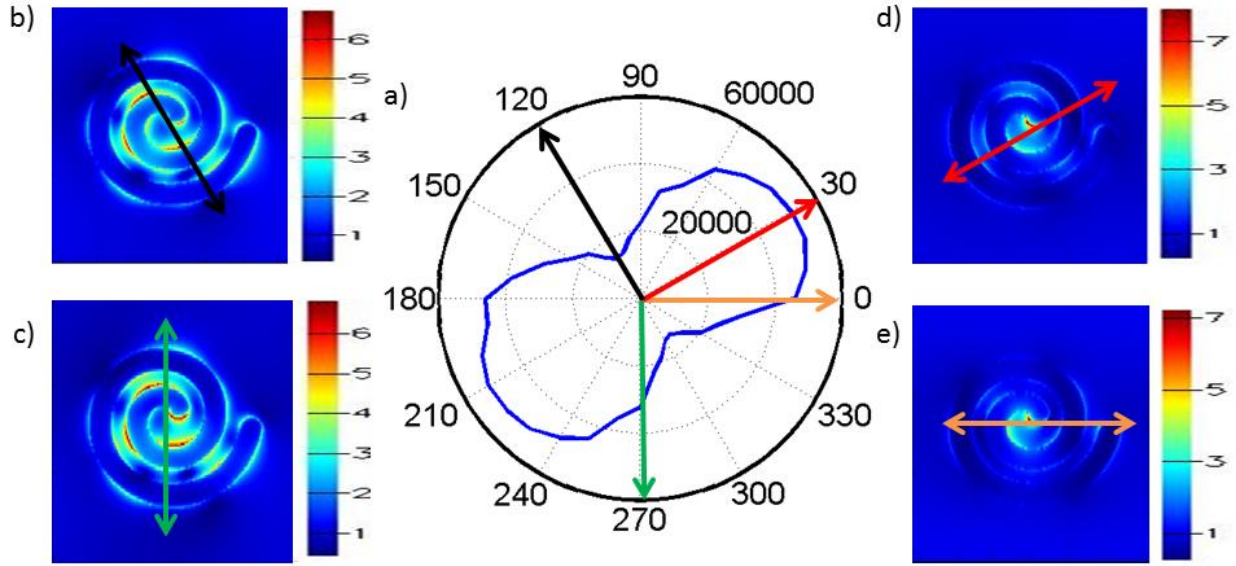

Figure 2 – a) SHG response to the axis of a linearly polarized excitation. 0 degrees corresponds to the line drawn between the beginning and end of the spiral (orange arrow). Maximum SHG signal occurs at 30 degrees (red arrow). Minimum SHG signal occurs at 120 degrees (black arrow). b-e) FDTD simulations of the electric-field strength for four different linear polarization excitation states

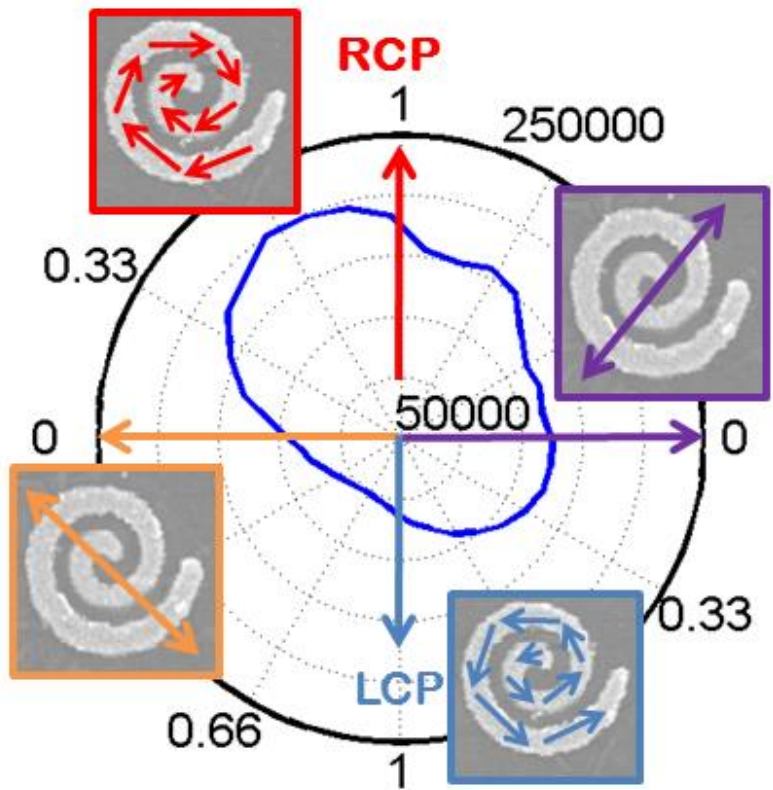

Figure 3 - SHG response to eccentricity and handedness of circular polarization. Horizontal axis corresponds to the two linear polarized states shown by the purple and orange lines. The vertical axis corresponds to completely circular polarizations.

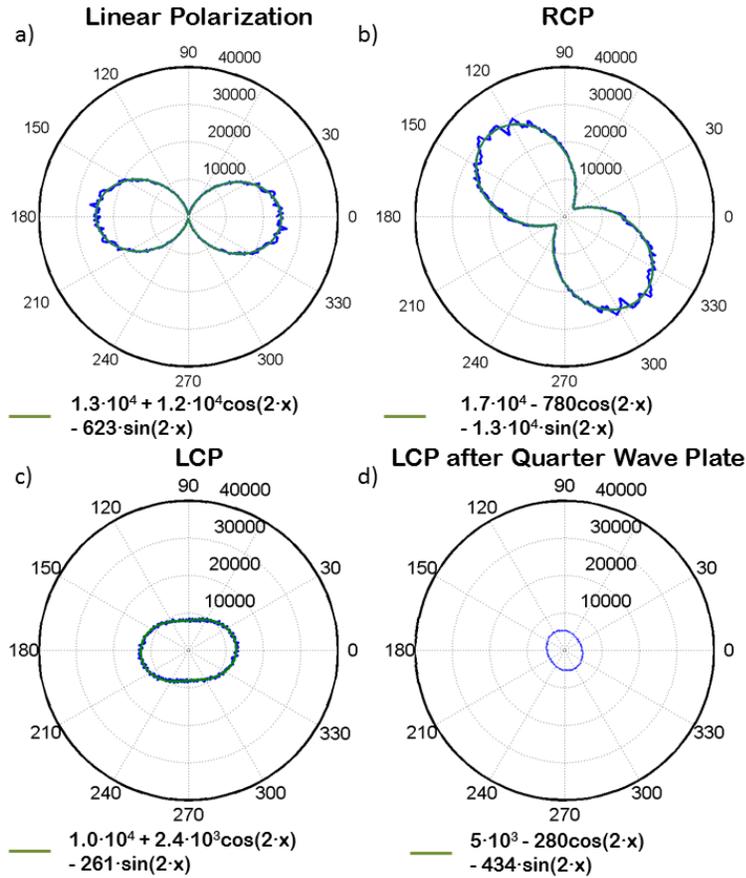

Figure 4 - Each graph shows SHG as a function of a linear analyzer placed after the nanoparticles a-c) was analyzed using only a linear polarizer. d) used a quarter wave plate before the linear analyzer. These measurements were fitted to a first order Fourier function to determine the polarization state of the emitted light. a) shows that linearly polarized incident light creates linear polarized harmonic light that was coaxial with the fundamental. b) shows that right handed circularly polarized incident light creates linear polarized harmonic light with a rotation of the polarization axis by 45 degrees. c) that left handed circularly polarized light creates harmonic light with no dominant axis of polarization. d) shows that the measurements taken in c) was a superposition of all polarizations.

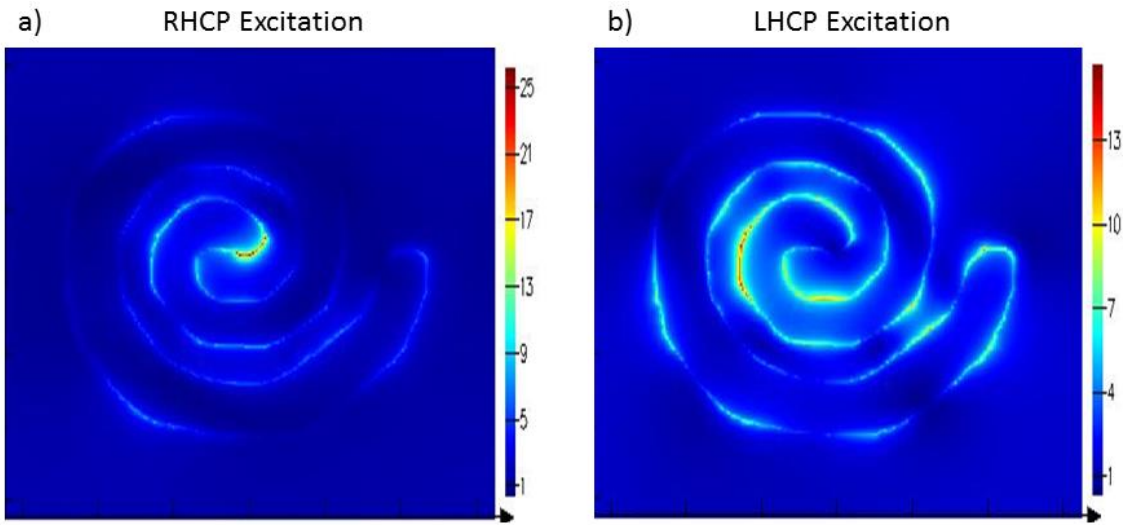

Figure 5 – FDTD simulations of the focusing mode of the nanospiral using a right handed circularly polarized excitation (a) and left handed circularly polarized light (b)

Supporting Information. More information on the efficiency calculations, saturation effects, and near field characteristics of the nanospiral is included. This material is available free of charge via the Internet at http://pubs.acs.org.

Corresponding Author

*Correspondence can be sent to: roderick.b.davidson@vanderbilt.edu

†Current address: Department of Molecular Biology and Genetics, Cornell University, Ithaca, New York

**Acknowledgements**. RBD, JIZ, GV and RFH received support from the Office of Science, United States Department of Energy (DE-FG02-01ER45916) for the lithographic fabrication of the nanospiral samples, the construction of the second-harmonic generation experiment and the measurements. SMA was supported by the National Science Foundation under a Phase II STTR grant to Kent Optronics (IIP-1058571) for the setup and maintenance of the laser source. The nanospiral samples were fabricated and characterized in facilities of the Vanderbilt Institute of Nanoscale Science and Engineering, which were renovated with funds provided by the National Science Foundation under the American Recovery and Reinvestment Act (NSF ARI-R2 DMR-0963361). A portion of this work was performed using EMSL, a national scientific user facility sponsored by the Department of Energy's Office of Biological and Environmental Research and located at Pacific Northwest National Laboratory. WPH and YG acknowledge support from the US Department of Energy, Office of Science, Office of Basic Energy Sciences, Division of Chemical Sciences, Geosciences & Biosciences.

References:

1.	Boyd, R. W., *Nonlinear optics*. Academic press: 2003.

2.	Grigorescu, A. E.; Hagen, C. W., Resists for sub-20-nm electron beam lithography with a focus on HSQ: state of the art. *Nanotechnology* **2009,** *20* (29), 292001.

3.	Aouani, H.; Navarro-Cia, M.; Rahmani, M.; Sidiropoulos, T. P. H.; Hong, M.; Oulton, R. F.; Maier, S. A., Multiresonant Broadband Optical Antennas As Efficient Tunable Nanosources of Second Harmonic Light. *Nano Letters* **2012,** *12* (9), 4997-5002.

4.	Czaplicki, R.; Zdanowicz, M.; Koskinen, K.; Laukkanen, J.; Kuittinen, M.; Kauranen, M., Dipole limit in second-harmonic generation from arrays of gold nanoparticles. *Opt. Express* **2011,** *19* (27), 26866-26871.

5.	Linden, S.; Niesler, F. B. P.; Förstner, J.; Grynko, Y.; Meier, T.; Wegener, M., Collective Effects in Second-Harmonic Generation from Split-Ring-Resonator Arrays. *Physical Review Letters* **2012,** *109* (1), 015502.

6.	Walsh, G. F.; Dal Negro, L., Enhanced second harmonic generation from Au nanoparticle arrays by femtosecond laser irradiation. *Nanoscale* **2013,** *5* (17), 7795-7799.

7.	Zhang, Y.; Grady, N. K.; Ayala-Orozco, C.; Halas, N. J., Three-Dimensional Nanostructures as Highly Efficient Generators of Second Harmonic Light. *Nano Letters* **2011,** *11* (12), 5519-5523.

8.	Eftekhari, F.; Davis, T. J., Strong chiral optical response from planar arrays of subwavelength metallic structures supporting surface plasmon resonances. *Physical Review B* **2012,** *86* (7), 075428.

9. Gorodetski, Y.; Drezet, A.; Genet, C.; Ebbesen, T. W., Generating Far-Field Orbital Angular Momenta from Near-Field Optical Chirality. *Physical Review Letters* **2013,** *110* (20), 203906.

10. Valev, V. K.; Baumberg, J. J.; Sibilia, C.; Verbiest, T., Chirality and Chiroptical Effects in Plasmonic Nanostructures: Fundamentals, Recent Progress, and Outlook. *Advanced Materials* **2013,** *25* (18), 2517-2534.

11. Valev, V. K.; Smisdom, N.; Silhanek, A. V.; De Clercq, B.; Gillijns, W.; Ameloot, M.; Moshchalkov, V. V.; Verbiest, T., Plasmonic Ratchet Wheels: Switching Circular Dichroism by Arranging Chiral Nanostructures. *Nano Letters* **2009,** *9* (11), 3945-3948.

12. Volkov, S. N.; Dolgaleva, K.; Boyd, R. W.; Jefimovs, K.; Turunen, J.; Svirko, Y.; Canfield, B. K.; Kauranen, M., Optical activity in diffraction from a planar array of achiral nanoparticles. *Physical Review A* **2009,** *79* (4).

13. Maier, S. A.; Atwater, H. A., Plasmonics: Localization and guiding of electromagnetic energy in metal/dielectric structures. *Journal of Applied Physics* **2005,** *98* (1), -.

14. Capretti, A.; Walsh, G. F.; Minissale, S.; Trevino, J.; Forestiere, C.; Miano, G.; Dal Negro, L., Multipolar second harmonic generation from planar arrays of Au nanoparticles. *Opt. Express* **2012,** *20* (14), 15797-15806.

15. Husu, H.; Siikanen, R.; Mäkitalo, J.; Lehtolahti, J.; Laukkanen, J.; Kuittinen, M.; Kauranen, M., Metamaterials with Tailored Nonlinear Optical Response. *Nano Letters* **2012,** *12* (2), 673-677.

16. Ziegler, J. I.; Haglund, R. F., Plasmonic Response of Nanoscale Spirals. *Nano Letters* **2010,** *10* (8), 3013-3018.

17. Ziegler, J. I.; Haglund, R. F., Complex Polarization Response in Plasmonic Nanospirals. *Plasmonics* **2013,** *8* (2), 571-579.

18. Pestov, D.; Lozovoy, V. V.; Dantus, M., Multiple Independent Comb Shaping (MICS): Phase-only generation of optical pulse sequences. *Opt. Express* **2009,** *17* (16), 14351-14361.

19. Weiner, A. M., Femtosecond pulse shaping using spatial light modulators. *Rev. Sci. Instrum.* **2000,** *71* (5), 1929-1960.

20. McMahon, M. D.; Lopez, R.; Haglund, R. F.; Ray, E. A.; Bunton, P. H., Second-harmonic generation from arrays of symmetric gold nanoparticles. *Physical Review B* **2006,** *73* (4).

21. Frank, B.; Yin, X.; Schäferling, M.; Zhao, J.; Hein, S. M.; Braun, P. V.; Giessen, H., Large-Area 3D Chiral Plasmonic Structures. *ACS Nano* **2013,** *7* (7), 6321-6329.

22. Schäferling, M.; Dregely, D.; Hentschel, M.; Giessen, H., Tailoring Enhanced Optical Chirality: Design Principles for Chiral Plasmonic Nanostructures. *Physical Review X* **2012,** *2* (3), 031010.

23. Papakostas, A.; Potts, A.; Bagnall, D. M.; Prosvirnin, S. L.; Coles, H. J.; Zheludev, N. I., Optical Manifestations of Planar Chirality. *Physical Review Letters* **2003,** *90* (10), 107404.

24. Kuwata-Gonokami, M.; Saito, N.; Ino, Y.; Kauranen, M.; Jefimovs, K.; Vallius, T.; Turunen, J.; Svirko, Y., Giant Optical Activity in Quasi-Two-Dimensional Planar Nanostructures. *Physical Review Letters* **2005,** *95* (22), 227401.

25. Ren, M.; Plum, E.; Xu, J.; Zheludev, N. I., Giant nonlinear optical activity in a plasmonic metamaterial. *Nat Commun* **2012,** *3*, 833.

26. Calander, N.; Gryczynski, I.; Gryczynski, Z., Interference of surface plasmon resonances causes enhanced depolarized light scattering from metal nanoparticles. *Chemical Physics Letters* **2007,** *434* (4–6), 326-330.

27. Heckel, J. C.; Chumanov, G., Depolarized Light Scattering From Single Silver Nanoparticles. *The Journal of Physical Chemistry C* **2011,** *115* (15), 7261-7269.

28. Kivshar, N. I. Z. Y. S., From metmaterials to metadevices. *Nature Materials* **2012,** (11), 7.